\pdfoutput=1
\documentclass{article}
\usepackage[utf8]{inputenc}
\usepackage[T1]{fontenc}
\usepackage{hyperref}
\usepackage{lmodern}
\usepackage{xspace}

\newcommand{\odebase}{ODEbase\xspace}

\begin{document}
\title{\odebase: A Repository of ODE Systems for Systems Biology}

\author{Christoph Lüders\\
University of Kassel, Germany\\
\href{mailto:chris@cfos.de}{\texttt{chris@cfos.de}}
\and
Thomas Sturm\\
CNRS, Inria, and the University of Lorraine, France\\
MPI Informatics and Saarland University, Germany\\
\href{mailto:thomas.sturm@loria.fr}{\texttt{thomas.sturm@loria.fr}}
\and
Ovidiu Radulescu\\
CNRS and the University of Montpellier, France\\
\href{mailto:ovidiu.radulescu@umontpellier.fr}{\texttt{ovidiu.radulescu@umontpellier.fr}}}

\date{January 2022}

\maketitle

\begin{abstract}
Recently, symbolic computation and computer algebra systems have been successfully applied in
systems biology, especially in chemical reaction network theory. One advantage of symbolic
computation is its potential for qualitative answers to biological questions. Qualitative methods
analyze dynamical input systems as formal objects, in contrast to investigating only part of the
state space, as is the case with numerical simulation. However, symbolic computation tools and
libraries have a different set of requirements for their input data than their numerical
counterparts. A common format used in mathematical modeling of biological processes is SBML. We
illustrate that the use of SBML data in symbolic computation requires significant pre-processing,
incorporating external biological and mathematical expertise.
\odebase provides high quality symbolic computation input data derived from established existing
biomodels, covering in particular the BioModels database.
\end{abstract}

\section{Introduction}

Recently, symbolic computation methods are playing an increasing role
in systems biology and mathematical biology
\cite[and the references there]{Boulier:18a}.
Problems investigated using such methods
include Hopf bifurcations, multi-stationarity,
multi-scale model reduction, dynamical invariants, and structural properties of steady state varieties;
for details see, e.g.,
\cite{Errami:15,gross:16,dickenstein:16,Bradford:20,kruff:21,grigoriev:20}.
Compared to numerical analysis and simulation, symbolic computation
provides not only quantitative, but qualitative results about
network dynamics, to some extent in parametric settings.
The biological systems investigated so far had a focus on reaction networks
in the sense of chemical reaction network theory \cite{Feinberg:19a}.
Such networks are usually
stored and exchanged in Systems Biology
Markup Language (SBML),
a free, open, and standardized XML-based format \cite{Hucka:03}.

On the one hand, symbolic computation does not utilize the full
information contained in SBML models.
For instance, SBML was designed with a focus on network simulation
and supports corresponding concepts like events and initial assignments,
which are not natural from a formal symbolic computation point of view.
On the other hand, symbolic computation
operates on formal objects, which are not readily available in SBML.
One prominent example are ordinary differential equations (ODEs)
describing differential network kinetics along with
algebraic constraints, such as conservation laws.
The genuine difference between dynamic simulation and
static formal analysis requires sensitivity to details
and rigor in the course of the construction of symbolic computation input from available SBML descriptions.
It is noteworthy that existing SBML parsers
generate input for numerical simulation, which is not suited
for symbolic computation.
MathSBML \cite{shapiro:04},
SBFC \cite{rodriguez:16},
SBMLtoODEpy \cite{ruggiero:19}, and
SBML2Modelica \cite{maggioli:20}
fall into that category.

It is important to understand that the rigorous construction
of symbolic computation input poses some substantial
problems.  Solving, or even recognizing, such problems,
requires joint competence and combined efforts not only from biology,
but also including mathematics and computer science.
We give some examples for such problems:

\begin{itemize}
  \item
  SBML allows floating-point values for various entities.
  However, floating-point values exhibit representation errors
  and computations are prone to rounding errors.
  This is inadequate for symbolic computation,
  where exact computations are performed.
  \item
  SBML has liberal naming conventions for species and parameters
  that interfere with the typically strict rules of symbolic computation
  software, which are oriented towards mathematical notation.
  If different users of symbolic computation software rename
  those identifiers at their own discretion, it becomes cumbersome
  to compare their results.
  \item
  SBML gives modelers versatile opportunities of expression,
  such as local parameters, function definitions,
  rules, and initial assignments.
  For practical reasons, scientific
  software does not generally support the
  full SBML feature set.  This leads to incompletely
  imported models or it prohibits the import entirely.
  \item
  Symbolic computation is concerned with mathematical properties
  like deficiency and linear conservation laws, which are
  available in SBML only implicitly through computation.
  Explicit availability is desirable, especially, since
  some of those computations can become surprisingly time-consuming.
\end{itemize}

With this in mind, the question is now how to make the
abundantly available SBML data accessible for the symbolic computation
community in a suitably prepared form.

A natural idea would be to integrate symbolic computation
input into the SBML format. However, there are obstacles on
both sides:
On the symbolic computation side, established software
is usually general-purpose, and systems biology is not yet
a strong focus of the community.
Therefore, widespread support
of SBML as an input format for symbolic computation software
cannot be expected in the near future.
On the systems biology side, the SBML standard would need
to be extended.
Standardization generally requires considerable efforts, and it seems
unlikely that this will be pursued before the links
between symbolic computation and systems biology
have been further strengthened.

The interdisciplinary project SYMBIONT brings together
researchers from mathematics, computer science, and systems biology \cite{Boulier:18a}.
Within SYMBIONT, we have started an online database
\emph{\odebase}, which
collects symbolic computation input for existing SBML models.
All models have been carefully constructed
taking into consideration the issues discussed above.
Models can be selected by biomathematical properties
such as deficiency, rank of the stoichiometric matrix, or
numbers of species, parameters, reactions, or conservation
constraints. Each model contains a link to its original SBML source.
At the time of writing, all our models
originate from the BioModels database \cite{lenovere:06},
the world's largest repository of curated mathematical
models of biological processes and one of the most important
data sources for modeling \cite{MalikSheriff:19}.
Out of the 1044 models from the curated branch of BioModels, we
have currently compiled 657 into \odebase.

As \odebase has turned out to be extremely valuable throughout the SYMBIONT project, we now make it
available to the community as a free and open database, beyond the lifespan of the
project.\footnote{\url{https://odebase.org}} If models require updates, revised versions will be
made available, keeping all previous versions for reference. Data can be extracted in Maple, Reduce,
SageMath, and \LaTeX{} format. We are open to supporting further formats in the future.

\odebase provides a canonical source of symbolic computation input
related to existing models of biological processes.
This has a number of advantages:
\begin{enumerate}
  \item
  Interdisciplinary competence:
  The derivation of adequate ODEs for the kinetics of
  existing biomodels requires to
  incorporate external biological and mathematical expertise.
  We have accomplished this task for a large set of available models
  and make the results available to the community.
  \item
  Economic use of human resources:
  Symbolic computation input has been pre-computed and is directly available.
  \item
  Availability:
  \odebase models used and cited in the literature
  can be conveniently reviewed on the basis of the original data
  and re-used in follow-up publications.
  \item
  Canonical reference:
  \odebase provides an unambiguous mapping of the, in general,
  too liberal SBML names for
  species concentrations and parameters to
  common mathematical notation.
  This facilitates comparability of results.
  \item
  Benchmarking:
  Beyond its primary purpose,
  ODEbase is perfectly suited to generate
  benchmark sets for novel algorithms and software
  in the field.
\end{enumerate}

\section{Details}

\subsection{The Content of \odebase Data Sets}
For each model in \odebase, the following data set is computed
from the original SBML input:

\paragraph{Stoichiometric and kinetic matrices.}
Stoichiometric and kinetic matrices are made explicit.
In that course, floating-point values in the SBML input
are converted to exact rational numbers.

\paragraph{ODEs for species concentrations.}
These are explicit first-order, non-linear ODEs that are often,
but not necessarily, autonomous.
Species are named $x_1$, \dots,~$x_n$, following common mathematical notation.
The ODEs are created from the stoichiometric matrix and
the relevant kinetic laws.
Again, floating-point values are
converted to exact rational numbers.
We have taken care to preserve the structure of mathematical
terms by using
abstract syntax trees as an intermediate format.  One visible
effect of this are uncanceled rational functions and the presence of
stoichiometric coefficients of $1$ or $-1$.
If species rate rules are present in the model,
the corresponding ODEs are included as well.

\paragraph{Parameter values.}
Our naming of parameters also follows common mathematical notation,
viz., $k_1$, \dots,~$k_m$.
Parameter values are converted from floating-point
representation to exact rational numbers.
If there are initial assignments or assignment rules in the SBML model,
they are applied in the proper order to calculate the parameter values.
To avoid any representation errors, all values
are queried as text from the XML source.

\paragraph{Map between \odebase names and original model names.}
A bijective mapping between the mathematical names for
species and parameters and their respective SBML names is provided.

\paragraph{Constraints.}
All SBML species assignment rules are converted to formal constraints.
Furthermore, linear conservation constraints are computed
from the stoichiometric matrix using an algorithm by Schuster and Höfer
\cite{Schuster:91},
extended to handle multiple model compartments.
All constraints introduced this way
can be combined with the ODEs mentioned above
which yields the relevant ODE system for the model.

\paragraph{Deficiency.}
The deficiency of the reaction network is computed from its complexes. This is a measure of how
independent the reaction vectors are, given the
network's linkage class structure
\cite[Sect.~6.3]{Feinberg:19a}.

\paragraph{Classification.}
The ODE right hand sides and both sides
of the constraints are analyzed for each model.
If all of these terms are polynomials,
the model is classified as \emph{polynomial}.
If all terms are rational functions, the model is
classified as \emph{rational}.

\paragraph{Test for mass-action kinetics.}
For all models we check whether the SBML-specified kinetics
differs from the regular mass-action kinetics
\cite[Sect.~2.1.2]{Feinberg:19a} only by a constant factor.
This is a conservative heuristic for identifying
models with mass-action kinetics.

\subsection{Supported SBML Features}
All models in \odebase are a faithful conversion from
the respective SBML model.
SBML features recognized during the conservation process include the following:
\begin{itemize}
  \item species with boundary condition,
  \item local parameters,
  \item parameter and species assignment rules,
  \item parameter and species initial assignments,
  \item species rate rules,
  \item function definitions.
\end{itemize}

SBML supports events, i.e., discrete model changes
at certain points in time, and furthermore it supports
parameter rate rules.
Models that contain either of those
are currently not included in \odebase.
Neither are models with irrational parameter values.

\section*{Acknowledgements}
Andreas Weber, who sadly passed away in 2020, originally gave
the idea for \odebase. Our student assistants
Anna Meschede and Matthias Neidhardt
at the University of Bonn supported us in
the programming the web front-end.

\section*{Funding}

This work has been supported by the interdisciplinary bilateral project
ANR-17-CE40-0036 and DFG-391322026 SYMBIONT
\cite{Boulier:18a}.


\providecommand{\noop}[1]{}

\end{document}